\documentclass[twocolumn,amsmath,amssymb,aps,pra,longbibliography,floatfix]{revtex4-1}

\usepackage{physics}
\usepackage{units}
\usepackage{amsmath}
\usepackage{cases}
\usepackage{url}
\usepackage{natbib}
\usepackage{textcase}
\usepackage{amssymb}
\usepackage{graphicx}
\usepackage{bm} 

\usepackage{times}
\usepackage{float}
\usepackage{multirow,microtype,color,relsize}
\usepackage{subfigure}
\usepackage[breaklinks=true]{hyperref}
\usepackage[utf8]{inputenc}
\usepackage[english]{babel}
\hypersetup{colorlinks=true,linkcolor=blue,citecolor=blue}
\hypersetup{linktocpage}
\usepackage{CJKutf8}
\usepackage{breakcites}
\usepackage{float}
\usepackage{siunitx}
\usepackage[dvipsnames]{xcolor}
\definecolor{mypine}{RGB}{1, 121, 111}

\begin{document}
\begin{CJK*}{UTF8}{gbsn}
\title{Metal-insulator transition in \texorpdfstring{$n$}{n}-type bulk crystals and films of strongly compensated SrTiO\texorpdfstring{$_3$}{3}}

\author{Yi Huang~(黄奕)}
\email[Corresponding author: ]{huan1756@umn.edu}
\author{Y. Ayino}
\author{B. I. Shklovskii}
\affiliation{School of Physics and Astronomy, University of Minnesota, Minneapolis, Minnesota 55455, USA}
\date{\today}

\begin{abstract}
We start by analyzing experimental data of Spinelli \textit{et al}~\cite{spinelli2010} for the conductivity of $n$-type bulk crystals of SrTiO$_3$ (STO) with broad electron concentration $n$ range of $4\times 10^{15}$ - $4 \times10^{20} $ cm$^{-3}$, at low temperatures. 
We obtain good fit of the conductivity data, $\sigma(n)$, by the Drude formula for $n \geq n_c \simeq 3 \times 10^{16} $ cm$^{-3}$ assuming that used for doping insulating STO bulk crystals are strongly compensated and the total concentration of background charged impurities is $N = 10^{19}$ cm$^{-3}$. 
At $n< n_c$, the conductivity collapses with decreasing $n$ and the Drude theory fit fails.
We argue that this is the metal-insulator transition (MIT) in spite of the very large Bohr radius of hydrogen-like donor state $a_B \simeq 700$ nm with which the Mott criterion of MIT for a weakly compensated semiconductor, $na_B^3 \simeq 0.02$, predicts $10^{5}$ times smaller $n_c$. 
We try to explain this discrepancy in the framework of the theory of the percolation MIT in a strongly compensated semiconductor with the same $N=10^{19}$ cm$^{-3}$. 
In the second part of this paper, we develop the percolation MIT theory for films of strongly compensated semiconductors. 
We apply this theory to doped STO films with thickness $d \leq 130$ nm and calculate the critical MIT concentration $n_c(d)$. 
We find that, for doped STO films on insulating STO bulk crystals, $n_c(d)$ grows with decreasing $d$. 
Remarkably, STO films in a low dielectric constant environment have the same $n_c(d)$. 
This happens due to the Rytova-Keldysh modification of a charge impurity potential which allows a larger number of the film charged impurities to contribute to the random potential. 
\end{abstract}

\maketitle
\end{CJK*}

\section{Introduction}
 SrTiO$_3$ (STO) is a classic example of perovskite oxides. 
 Historically, the primary interest in this material was on its unusual dielectric properties. 
 The dielectric constant $\kappa$ increases from around 300 at $T=300$ K to 20000 at low temperatures. 
 Due to its widespread use as a single-crystal substrate for epitaxial growth of oxides, its commercial availability, controllable surface termination, and close lattice match to numerous complex oxide materials, STO has been employed as an active component in oxide heterostructures and thin films. 
 Examples include, its use as a dielectric layer in field-effect devices for tuning the carrier density of materials such as cuprates, and in such systems as $\text{SrTiO}_3 /\text{LaAlO}_3$ forming a two-dimensional electron gas interface. 

In this paper, we are concerned with fascinating and potentially useful, electronic transport properties of STO, which is a semiconductor with a 3.2 eV gap.
$n$-type conduction of STO has been most commonly achieved by substitution of $\text{La}^{3+}$ for $\text{Sr}^{2+}$, $\text{Nb}^{5+}$ for $\text{Ti}^{4+}$, or by oxygen reduction to $\text{SrTiO}_{3-\delta}$. 
Similarly, $p$-type doping has been achieved by substituting trivalent metal ions, such as $\text{In}^{3+}$, $\text{Al}^{3+}$, $\text{Fe}^{3+}$, and $\text{Sc}^{3+}$ for $\text{Ti}^{4+}$. 
Transport data for electron doped STO bulk crystals are summarized in Ref.~\cite{spinelli2010}. 
Within the hydrogenic theory of shallow donors, the donor Bohr radius, $a_B = \hbar^{2} \kappa/m^{\star}e^{2} = 0.053 \kappa m_e/m^{\star}$ nm, where $m^{\star} \simeq 1.4 m_e$ \cite{uwe1985, allen2013} is the effective electron mass and $m_e$ is free electron mass. 
At $T=4$ K, $a_B \simeq 700$ nm is so large that the Mott criterion for the metal-insulator transition (MIT) in doped weakly compensated semiconductors~\cite{mott1969}, $n_c a_B^3 \simeq 0.02$, leads to a very small $n_c \simeq 5 \times 10^{10} $ cm$^{-3}$. 
Thus, at all experimental concentrations of donors and electrons $n > 10^{15} $ cm$^{-3}$, STO should be strongly degenerate and should have metallic conductivity. 

Reference~\cite{spinelli2010} presents temperature dependencies of resistivity for a large set of $n$-type STO samples with electron concentrations $n$ ranging from $3.8 \times 10^{15}$ to $3.79 \times 10^{20} $ cm$^{-3}$ . 
In Fig.~\ref{fig:sigma}, we plot by red dots the conductivity at 4 K obtained in Ref.~\cite{spinelli2010} as a function of $n$. 
(All the lines at Fig. \ref{fig:sigma} are theoretical results and will be explained later). 
We see that the character of $n$ dependence of $\sigma$ dramatically changes around $n_c \simeq 3 \times10^{16} $ cm$^{-3}$.  
This apparent MIT at $n_c \simeq 3 \times10^{16}$ cm$^{-3}$ is in dramatic contradiction with the theoretical prediction, based on the Mott criterion. 

In this paper, we address this puzzle and show that the Mott criterion fails because insulating bulk crystals of STO used for intentional doping are almost completely compensated, i.e., have large and almost equal concentrations $N/2$ of background charged donors and acceptors. 
When the intentionally added concentration of donors and electrons $n$ is much smaller than $N$, MIT is driven by the random long-range Coulomb potential of $N$ charged donors and acceptors~\cite{shklovskii1971,shklovskii2013}. 
With decreasing $n$, the Fermi energy of degenerate electron gas gets smaller while the screening of Coulomb potential of impurities gets weaker. 
At some $n = n_c$, the amplitude of potential fluctuations becomes larger than the depth of the electron Fermi sea and electrons get localized in large puddles separated from each other by large potential barriers. 
We call such a transition the percolation MIT in strongly compensated semiconductor, to discriminate it from the Mott MIT in weakly compensated semiconductor.

The plan of this paper is as follows. 
In Sec. II, we analyze the experimental data of Spinelli \textit{et al}~\cite{spinelli2010} for conductivity at 4 K shown in Fig.~\ref{fig:sigma}.  
We calculate the Drude conductivity, $\sigma(n,N)$, of compensated STO assuming that the scattering happens only on randomly positioned Coulomb impurities. 
For $n \geq n_c \simeq 3 \times 10^{16} $ cm$^{-3}$, we obtained reasonably good fit of the conductivity data, $\sigma(n)$, with the concentration of background charged impurities $N = 10^{19}$ cm$^{-3}$. 
Using this $N$, we show that the theory of the percolation MIT of Refs.~\cite{shklovskii1971,shklovskii2013} gives $n_c$ close to the experimental value.

In Sec. III, we extend the percolation MIT theory to films of strongly compensated STO with thickness $d \leq 130$ nm and calculate the critical MIT electron concentration $n_c(d)$. 
We find that $n_c(d)$ grows with decreasing $d$, because the Rytova-Keldysh modification of the Coulomb potential of a charge impurity slows down its potential decay within the STO film, and allows a larger number of the film impurities to contribute to the random potential breaking Fermi sea in puddles. 
The theory of the percolation MIT developed here for STO films is valid for all other compensated semiconductor films.

\section{Conductivity and metal-insulator transition in STO bulk crystals.}

Let us start from the low temperature theory of metallic conductivity. At large $n$, when $na_B^3 \gg 1$, electron gas is degenerate and  $k_F l \gg1$, where $k_F = (3\pi^2 n)^{1/3}$ is the Fermi wave vector and $l$ is the mean free path, so that one can use the Drude formula for conductivity: $\sigma = n e^2 \tau/m^{\star}$,
where $\tau = l/v_F$ is the momentum relaxation time, $v_F = \hbar k_F/m^{\star}$ is the Fermi velocity, and $e$ is the charge of an electron. 
At low temperatures the dominant scattering mechanism is due to ionized donors. 
In the standard Thomas-Fermi approximation, the screening radius of the degenerate electron gas~\cite{mott1936,mansfield1956}
$r_s= [(a_B/4)(\pi/3n)^{1/3}]^{1/2}$ is much larger than the electron wavelength, $k_F r_s \gg 1$, so one can use the screened Coulomb potential $\phi(r) = (e/r)e^{-r/r_s}$ to compute the momentum relaxation time and the corresponding conductivity.
Under the assumption that compensation is absent and the number of ionized impurity centers is the same as the number of free carriers, $n$, we get for the conductivity of a degenerate gas
\begin{equation}
    \sigma_1 = \frac{e^2}{\hbar a_B} \frac{3 \pi}{2} n a_B^3 f[(3\pi^5 n a_B^3)^{1/3}], \label{eq:sigma10}
\end{equation}
where $f(x) = [\ln(1+x) - x/(1+x)]^{-1}$.

\begin{figure}[t]
    \includegraphics[width=\linewidth]{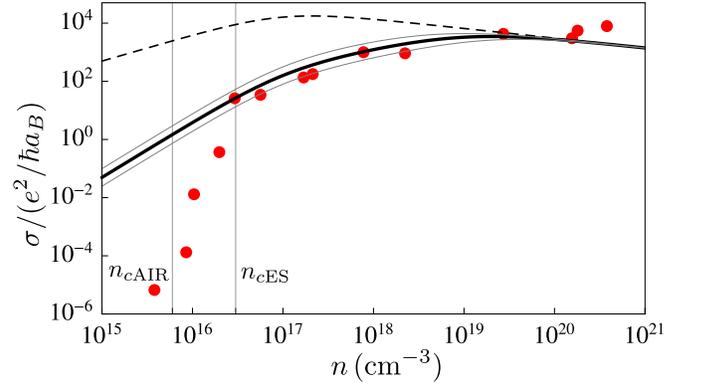}
    \caption{Dimensionless conductivity $\sigma / (e^2/\hbar a_B)$ as a function of carriers concentration $n$ at 4 K. Experimental data~\cite{spinelli2010} are shown by red dots. 
    All data for $n < 3\times 10^{17}$ cm$^{-3}$ are obtained by reduction of insulating bulk crystals, while higher concentration sample are mostly doped by Nb. Theoretical predictions based on interpolation between Eqs. (\ref{eq:sigma1}) and (\ref{eq:sigma2}) are shown by full line for strongly compensated STO with the background impurity concentration  $N= 10^{19} $ cm$^{-3}$, and by dashed line for uncompensated STO using Eqs.~\eqref{eq:sigma10} and \eqref{eq:sigma20}. Upper thin gray line shows how results change for $N = 5 \times 10^{18} $ cm$^{-3}$, while lower thin gray line is for $N = 2\times 10^{19} $ cm$^{-3}$. Critical concentrations of MIT are shown by two thin vertical lines as predicted by Anderson-Ioffe-Regel theory (left) and Efros-Shklovskii theory (right) for $N = 10^{19} \text{cm}^{-3}$.}
\label{fig:sigma}
\end{figure}

Due to the large dielectric constant, the scattering cross section, $\Sigma \propto n^{-4/3}$, used to calculate Eq. (\ref{eq:sigma10}), gets very small quickly, as $n$ increases. 
For $n \simeq 10^{17} $ cm$^{-3}$, $\Sigma \simeq a^2$, where $a \simeq 4$ $\si{\angstrom}$ is the lattice constant for STO. 
Following Ref. ~\cite{wemple1965}, we argue that once $\Sigma \simeq a^2$, it saturates and can not get any smaller with increasing $n$. 
Indeed, for length scales of the order $a$, there is no more dielectric screening, so the electron ``feels'' the full potential of the charged impurity center $e/a$. 
It leads to the geometrical cross section of the donor of the order of $a^2$.
As a result, for $n > n_{\rm core}  \sim 10^{17} $ cm$^{-3} $, Eq. (\ref{eq:sigma10}) is no longer valid and the conductivity is given by
\begin{align}
\sigma_2  = \frac{e^2}{\hbar a_B} \frac{a_B^2}{a^{2}} (3\pi^2 n a_B^3)^{-1/3}.\label{eq:sigma20}
\end{align}
We see from Eq. (\ref{eq:sigma20}) that $\sigma(n)$ decreases with increasing $n$ as  $n^{-1/3}$. This dependence was first derived in Refs.~\onlinecite{morita1963,shimizu1963} and observed in heavily doped PbTe and  SnTe in Ref.~\onlinecite{allgaier1962}.  
Using the  Matthiessen addition rule, $\sigma^{-1} = \sigma_1^{-1} + \sigma_2^{-1}$, to interpolate between Eqs. (\ref{eq:sigma10}) and (\ref{eq:sigma20}), we plot the dimensionless conductivity $\sigma/(e^2/\hbar a_B)$ versus the carriers concentration $n$ in Fig.~\ref{fig:sigma} by the dashed line. 
We see that at relatively small $n$ such ignoring compensation theory predicts 2-3 orders of magnitude larger conductivity than the data. 

We argue that the experimental conductivity is small because of almost complete compensation of insulating STO crystals used in Ref.~\onlinecite{spinelli2010} and universally for intentional doping by donors. Namely, these insulating samples contain uncontrolled and practically equal concentrations of background donors ($N_D$) and  acceptors ($N_A$)~\cite{tufte1967,spinelli2010,Ambwani2016}. 
Thus, the total concentration of charged impurities $N = N_A+N_D$ is large and this strongly reduces the conductivity. Below, we present an evidence that $N \sim 10^{19}$ cm$^{-3}$. 

Background impurities increase the total number of Coulomb impurities to $(n+N)$, and replace Eqs.~\eqref{eq:sigma10} and \eqref{eq:sigma20} by the following two equations:
\begin{align}
    \sigma &= \frac{e^2}{\hbar a_B} \frac{3 \pi}{2} \frac{n}{n+N} n a_B^3 f[(3\pi^5 n a_B^3)^{1/3}]\qc n<n_{\rm core},  \label{eq:sigma1}\\
    \sigma &=  \frac{e^2}{\hbar a_B} \frac{a_B^2}{a^{2}} \frac{n}{n+N}(3\pi^2 n a_B^3)^{-1/3}\qc n>n_{\rm core}. \label{eq:sigma2}
\end{align}

We found that conductivity of a strongly compensated sample interpolated between Eqs. (\ref{eq:sigma1}) and (\ref{eq:sigma2}) gives the best fit to the experimental data. This fit is shown in Fig.~\ref{fig:sigma} by the thick black line at $N = 10^{19}$ cm$^{-3}$, which agrees much better with the data than the dashed line obtained for uncompensated samples. 
However, for larger $n > N$, while theory predicts that conductivity should go down with $n$ as $\sigma \propto n^{-1/3}$, the experiment shows that it saturates or even slightly (like $n^{0.1}$) grows. This remains an unsolved puzzle~\footnote{In the recent paper~\cite{verma2014} the authors used an unrealistic parameters such as $ \kappa  =  4$ instead of $\simeq 20,000$ and the effective mass $m^{\star} \simeq 6.5 m_e$ instead of $1.4m_e$ in order to fit data of Ref. \cite{cain2013}. In this way they were able to get inequality $k_F r_s \ll 1$, which allowed them to use theory of Ref.~\cite{frederikse1967} to predict a saturating behavior for conductivity at large $n$ and avoid dealing with compensation. However, as the authors admit there is no justification for using such parameters. In reality, $k_F r_s \gg 1$ and the theory~\cite{mott1936,mansfield1956} is applicable.}.

Mott's condition for MIT critical concentration, $n_c a_B^3 \simeq 0.02$, was derived for weakly compensated samples, so it is understandable that it does not predict the correct critical concentration for strongly compensated samples. 
To include the effect of compensation one can try the Anderson-Ioffe-Regel (AIR) condition $k_{F}l \simeq 1$ for MIT\cite{anderson1958,ioffe1960}. 
The mean free path of charge carriers in a strongly compensated STO sample with $N \gg n$ is  
\begin{equation} 
    l = a_B \frac{(3\pi^2 n a_B^3)^{4/3}}{2 \pi N a_B^3} f[(3\pi^5 n a_B^3)^{1/3}]  \label{eq:mean_free_path}
\end{equation}

Assuming that $N = 10^{19} $ cm$^{-3}$, we find the critical concentration for MIT to be $n_{c {\rm AIR}} \simeq 6 \times 10^{15} $ cm$^{-3}$. 
However, as shown in Fig.~\ref{fig:sigma}, the predicted $n_{c {\rm AIR}}$ is still five times smaller than experimental $n_c$. 
To understand such discrepancy, notice that the Anderson-Ioffe-Regel MIT criterion is justified for scattering by individual random charges calculated in Born approximation.
However, Efros and Shklovskii~\cite{shklovskii1971,shklovskii2013} showed that the physics of strongly compensated semiconductor is dominated by large long range Coulomb potential of large number of charged impurities which, because of week screening at MIT, becomes larger than the Fermi energy of electrons and, therefore, breaks the electron gas in puddles separated by potential barriers at $n=n_{c {\rm ES}}$. 
Clearly, such situation can not be dealt with in Born approximation. 
This transition happens while locally, in puddles, electron gas is still a good metal with $k_{F}l \gg 1$. 
This means that $n_{c {\rm ES}} \gg n_{c {\rm AIR}}$. 
Predicted~\cite{shklovskii1971,shklovskii2013} $n_{c {\rm ES}}$ is given by
\begin{align}
n_{c {\rm ES}} = \beta \frac{N^{2/3}}{a_B}, \label{eq:nc_bulk}
\end{align}
where $\beta$ is a numerical coefficient. 
Analysis of experiments in compensated germanium shows~\cite{shklovskii2013} that $\beta \simeq 0.5$. 
Using $N = 10^{19}$ cm$^{-3}$, we find that $n_{c {\rm ES}} \simeq 3 \times 10^{16} $ cm$^{-3}$.  
We see from the Fig.~\ref{fig:sigma} that for strongly compensated semiconductor this critical concentration agrees much better with experimental MIT than $n_{c {\rm AIR}}$. 
As we mentioned in Introduction the Mott criterion for weakly compensated semiconductor (where it is equivalent to AIR),  $n_c a_B^3 \simeq 0.02$, is more than $10^{5}$ times off.

The derivation of Eq. (\ref{eq:nc_bulk}) is very simple~\cite{shklovskii1971,shklovskii2013}. 
One estimates the amplitude of fluctuations of the bottom of conduction band in electrostatic potential fluctuations $\gamma \sim (e^{2}/\kappa r_s)(Nr_s^{3})^{1/2}$ in a good metal with randomly distributed Coulomb impurities, where $r_s(n)$ is the screening radius for a degenerate electron gas. 
Then, one equates $\gamma$ to the Fermi energy $\hbar^{2} (3\pi^2 n)^{2/3}/2m$ and solves for $n_{c {\rm ES}}$.

Although Eq. (\ref{eq:nc_bulk}) agrees with apparent MIT at $n_{c} \simeq 3 \times 10^{16} $ cm$^{-3}$ our theory meets two big problems when compared to the data~\cite{spinelli2010}.

First, one should recall that this is a zero temperature theory. 
It can work at finite temperature only if at $n=n_c$ both the Fermi energy and the amplitude of disorder potential $\gamma(n_c)$ are larger than $k_{B}T$. 
In reality simple estimates show that at temperature 4 K they are four times smaller than $k_{B}T$. 
Thus, no transition and localization of electrons in puddles can happen at $T=4$ K. 
At $n < n_c$, energy $\gamma(n)\sim e^2N^{2/3}/n^{1/3}$, it grows with decreasing $n$, but quite slowly~\cite{shklovskii1971,shklovskii2013}. 
Even for the sample with $n = 8.5\times 10^{15}$ cm$^{-3}$ $\gamma(n)$ is still 2 times smaller than $k_{B}T$ at $T=4$K. 
How can this sample have $\sigma(n) \sim 10^{-6} \sigma(n_c)$? 
This may happen only if the disorder potential is 30 times larger than our $\gamma(n)$. 
We have no idea how to explain such apparent large energy scale of disorder. 
Even if we were able to do that, we would arrive much larger $n_c$ in disagreement with experimental data~~\cite{spinelli2010}.

Second, the resistivity of the sample with $n = 8.5 \times 10^{15}$ cm$^{-3}$ measured down to $T=1$ K while being very large showed apparent metal-like saturation at low temperatures. 
This disagrees with expected in strongly compensated semiconductor low temperature variable range hopping conductivity~\cite{shklovskii1973}. 
We have no explanation for such behavior.

It was assumed above that donors and acceptors positions are random, i.e., they are not making compact donor-acceptor pairs. 
For samples made by cooling from a melt, the distribution of impurities in space is a snapshot of the distribution that impurities have at higher temperature, when their diffusion practically freezes~\cite{keldysh1964}. 
In relatively narrow band gap semiconductors, at this temperature there is a concentration of intrinsic carriers larger than the concentration of impurities. 
Intrinsic carriers thus screen the Coulomb attraction between donor and acceptors, so that impurities remain randomly distributed in space. 
As a result, when the temperature is lowered to the point where intrinsic carriers recombine, the impurities are left in random positions~\cite{gal'pern1972,shklovskii2013}.
STO is a wide gap semiconductor so that explanation based on intrinsic carriers does not work. 
However, at temperatures of diffusion freezing, say 1000 K, STO still has a very large dielectric constant $\kappa \sim 100$. 
Therefore, acceptors and donors of strongly compensated STO can avoid making pairs~\footnote{We can use analogy with water solutions ($\kappa=81$) of simple salt such as NaCl, where ions $\text{Na}^+$ and $\text{Cl}^-$ play the role of our donors and acceptors. It is known that at $T=300$ K salt stays dissociated till concentrations as large as $10^{21} $ cm$^{-3}$. Here we are dealing with similar $\kappa$, comparable to two hydration radii of ions in water minimum distance between donor and acceptor (lattice constant of STO), but larger $T$ and much smaller concentration $N \simeq 10^{19} $ cm$^{-3}$. Thus, almost all donor-acceptor pairs should be dissociated at temperature of diffusion freezing in STO.}.

\section{Metal-insulator transition in STO films}

STO films attract growing attention~\cite{kozuka2009,Kozuka2010,verma2014,Ambwani2016,Liu2018}. 
They are made by three different methods: molecular beam epitaxy (MBE), pulse laser deposition (PLD), and spattering. 
MBE films are grown from high purity elements and may have less impurities and are relatively clean. 
PLD films are often grown from STO single crystal targets and are presumed to have the same or close composition, i.e., they are strongly compensated and have approximately the same $N$ as STO bulk crystals. 
Spattered films are typically grown from polycrystalline bulk targets and have even more impurities~\cite{Ambwani2016}. 

In this paper, we focus on PLD films. 
They can be intentionally doped by donors and studied in large range of concentration of electrons $n$. 
In this section, we calculate the critical concentration of electrons at MIT $n_c(d)$ as a function of the film thickness $d$. 
We show below that because of very large dielectric constant of STO, the dielectric constant of the film environment dramatically affects the amplitude of the long range random potential of Coulomb impurities of the film and, therefore, the critical concentration of MIT $n_c(d)$. 
Here we consider two interesting experimental situations: $n$-type doped STO film surrounded by materials with much smaller dielectric constant $\kappa_e$ and doped STO thin film on insulating STO substrate. 

\subsection{STO film in low dielectric constant environment}

In this section, we study thin STO films with dielectric constant $\kappa$ in environment with dielectric constant $\kappa_e\ll\kappa$,  for example STO grown on (and capped with) a non-STO perovskite insulators with much lower dielectric constants $\kappa_e \sim 25$. In this case, according to Rytova and Keldysh~\cite{rytova1967,keldysh1979,cudazzo2011}, the electric field lines of a charged impurity channel through the thin film before exiting outside to the environment at distance $r_0 = \kappa d/2\kappa_e$.
At $d < r < r_0$, the potential of such impurity is only logarithmically different from $e^2/\kappa d$. 
Electrons screen this potential at some screening radius $\lambda(n)$. 
At MIT, the screening becomes nonlinear and $\lambda$ can be estimated from the condition that fluctuations of impurity charge concentration in the volume $d \lambda^2$ can be barely compensated by the redistribution of the concentration of electrons $n$, such that $n \sim (Nd \lambda^2)^{1/2}/d\lambda^2$. 
This gives
\begin{equation}\label{eq:lambda}
 \lambda \sim \frac{1}{n}\qty(\frac{N}{d})^{1/2}.
 \end{equation}
The corresponding amplitude of the Coulomb potential energy fluctuations is 
\begin{equation}\label{eq:gamma}
    \gamma \sim \frac{e^2}{\kappa d} (Nd \lambda^2)^{1/2} \sim \frac{e^2 N}{\kappa n d}.
\end{equation}

In order to estimate the critical concentration of the MIT, we equate the potential energy to the Fermi energy
\begin{equation}\label{eq:critical_mu}
    \gamma \sim \hbar^2 k_F^2/2m^{\star}.
\end{equation}
To proceed further, we should relate the local Fermi wave vector $k_F$ with the local density $n$.
First, consider a very thin film such that the motion along the perpendicular to film $z$ direction is quantized and restricted to the lowest subband.
Namely, the Fermi energy is smaller than the subband gap $\sim \hbar^2/m^{\star} d^2$, or equivalently $d \lesssim k_F^{-1}$.
In such a thin film, electrons have only two degrees of freedom parallel to the film plane, and $k_F \sim \sqrt{n d}$.
Substituting $k_F \sim \sqrt{n d}$ back into Eq.~\eqref{eq:critical_mu}, we arrive at the critical concentration
\begin{equation}\label{eq:nc_keldysh1}
    n_c(d) \sim \frac{1}{d} \qty(\frac{N}{a_B})^{1/2}\qc d \lesssim d_1,
\end{equation}
where $a_B = \kappa\hbar^2/m^{\star}e^2$ is the effective STO Bohr radius.
The above result is self-consistent if $d \lesssim k_F^{-1} \sim (n d)^{-1/2}$, which gives $d \lesssim d_1 =a_B/(N a_B^3)^{1/4}$.
For $N\simeq 10^{19}$ cm$^{-3}$ and $a_B \simeq 700$ nm, we have $d_1 \simeq 15$ nm.

\begin{figure}[t]
    \centering
    \includegraphics[width = \linewidth]{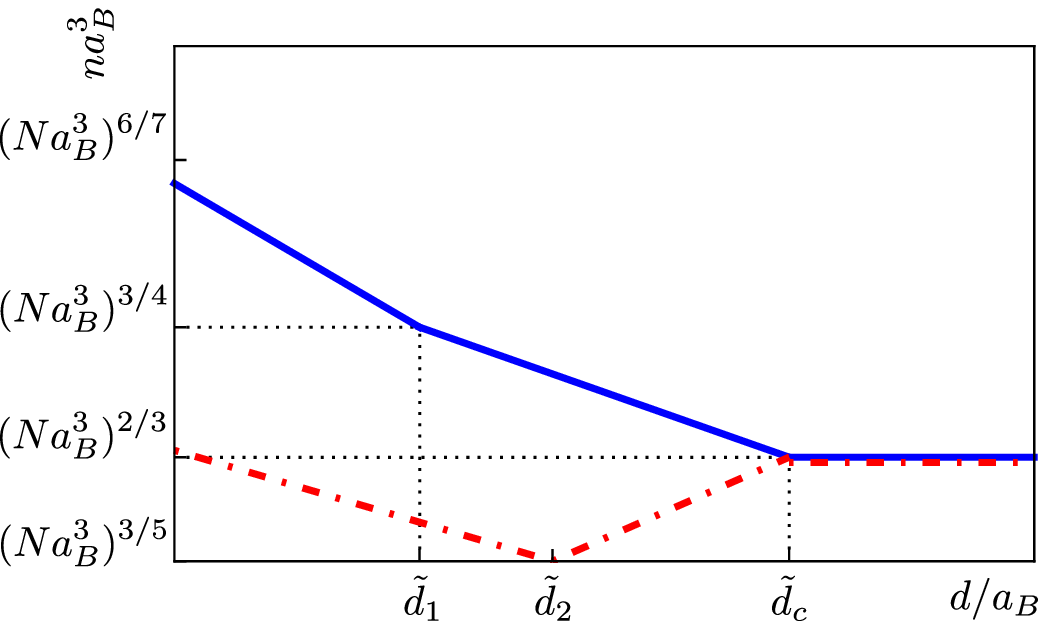}
    \caption{The phase diagram of MIT in the plane $(n,d)$ plotted in log-log scale using dimensionless variables $na_B^3$ and film thickness $\tilde{d} = d/a_B$. The blue solid curve corresponds to phase boundary Eqs.~\eqref{eq:nc_keldysh1}, \eqref{eq:nc_keldysh2}, and \eqref{eq:nc_bulk} for doped STO film both in low dielectric constant environment (say, silicon oxide) and on insulating STO substrate.
    The red dot-dashed curve corresponds to Eqs.~\eqref{eq:nc_normal1}, \eqref{eq:nc_normal2} and \eqref{eq:nc_bulk} in the case of doped STO film on MBE STO substrate. Above each curve the system is metallic, while below the curve it is insulating. The dimensionless thickness tick marks are $\tilde{d}_1 = (Na_B^3)^{-1/4}$, $\tilde{d}_2 = (Na_B^3)^{-1/5}$, and $\tilde{d}_c = (Na_B^3)^{-1/9}$.}
    \label{fig:nc_film}
\end{figure}

Next, consider a thicker film with $d \gtrsim k_F^{-1}$ and $k_F \sim n^{1/3}$. In this case, Eq.~\eqref{eq:critical_mu} gives the critical concentration
\begin{equation}\label{eq:nc_keldysh2}
    n_c(d) \sim \qty(\frac{N}{a_B d})^{3/5}\qc d_1 \lesssim d \lesssim d_c,
\end{equation}
At $d \sim d_1$, Eq.~\eqref{eq:nc_keldysh2} crosses over to Eq.~\eqref{eq:nc_keldysh1}, while at $d \gtrsim d_c = a_B/(N a_B^3)^{1/9}$, Eq.~\eqref{eq:nc_keldysh2} matches the bulk value Eq. (\ref{eq:nc_bulk}). 
For $N\simeq 10^{19}$ cm$^{-3}$ and $a_B \simeq 700$ nm, we have $d_c \simeq 130$ nm.
Eqs.~\eqref{eq:nc_keldysh1}, \eqref{eq:nc_keldysh2}, and \eqref{eq:nc_bulk} are shown by the blue solid curve in Fig.~\ref{fig:nc_film}.

Since the above results Eqs.~\eqref{eq:nc_keldysh1} and \eqref{eq:nc_keldysh2} are based on $d < \lambda < r_0$, one should check if it as indeed satisfied near our MIT line. 
Inequality $d < \lambda < r_0$ gives the restriction on $n$
\begin{equation}\label{eq:n3_bound}
    \frac{\sqrt{N}}{d^{3/2}} \frac{\kappa_e}{\kappa} < n < \frac{\sqrt{N}}{d^{3/2}}.
\end{equation}
The upper and lower bounds on $n$ Eq.~\eqref{eq:n3_bound} cross the curve $n_c(d)$ Eqs.~\eqref{eq:nc_keldysh1} and \eqref{eq:nc_keldysh2} at $d \sim d_0 = a_B (\kappa_e/\kappa)^2$ and $d \sim d_c$ respectively.  
For typical $\kappa/\kappa_e \sim 10^3$, the lower limit $d_0 \ll 1$ nm. 
Thus, our phase boundary $n_c(d)$ is valid at all reasonable film widths $d$.

Above we ignored the concentration of charged impurities in the environment of the STO film, $N_e$. 
Let us now evaluate the role of such impurities following similar analysis in the case of topological insulator films~\cite{huang2021}. 
To save the electrostatic energy, the electric field lines of an impurity at distance $z \lesssim r_0$ from the film surface first enter inside the STO film and then radially spread inside the film to distance $\sim r_0$ before exiting outside the film to infinity. 
Thus one can think that effectively each outside impurity is represented inside the film by a charge $e$ disk, with radius $z$ and thickness $d$. 
In the presence of screening, only small minority of the outside impurities with $z < \lambda$ contribute in fluctuating charge of the volume $d\lambda^2$. 
As a result, total effective concentration of impurities projected from outside the film is $N_e\lambda/d$.
If $N_e\lambda_c/d< N$, where $\lambda_c$ is given by Eq.~\eqref{eq:lambda} at $n=n_c(d)$, outside impurities can be ignored and our results for $n_c(d)$ are valid.
Using  Eqs.~\eqref{eq:nc_keldysh1} and \eqref{eq:nc_keldysh2}, we get $\lambda_c/d =(a_B/d)^{1/2}$ at $d< d_1$ and  $\lambda_c/d=(a_B/d)^{9/10}(Na_B^{3})^{-1/10}$ at $d_1 < d < d_c$.
We see that $\lambda_c/d$ grows with decreasing $d$, making small $d$ more vulnerable to external impurities. 
Thus, for $d<d_1$, the condition of validity of above results $N_e\lambda_c/d< N$ is $d > a_B (N_e/N)^{2}$.

\subsection{Doped STO film on the insulating STO substrate} 

In this case, the dielectric constant is uniform and we can deal with the conventional Coulomb potential of charged impurities with uniform concentration $N$. 
At MIT, the screening by the electrons of doped STO film becomes nonlinear. 
The nonlinear screening radius $\lambda$ can be estimated from the condition that fluctuations of the impurity charge concentration in the volume $\lambda^3$ of a cube including a square $\lambda^2$ of the STO film can be barely compensated by redistribution of the concentration of electrons $n$ inside the film~\cite{Shklovskii1986}: 
\begin{equation}\label{eq:n3_uniform}
    nd \sim (N\lambda^3)^{1/2}/\lambda^{2}.
\end{equation}
This gives
\begin{equation}\label{eq:lambda1}
 \lambda \sim \frac{N}{n^2d^2}.
\end{equation}
The corresponding amplitude of the Coulomb potential fluctuations therefore is
\begin{equation}\label{eq:gamma1}
    \gamma \sim \frac{e^2}{\kappa \lambda} (N \lambda^{3})^{1/2} \sim \frac{e^2 N}{\kappa n d}.
\end{equation}
Remarkably, we arrive at the same result Eq.~\eqref{eq:gamma}, as for doped STO films in a low dielectric constant environment. 
This means that for the case of a doped STO film on an insulating STO substrate, the phase boundary $n_c(d)$ is identical to one given by Eqs.~\eqref{eq:nc_keldysh1} and \eqref{eq:nc_keldysh2} and shown by the blue line of Fig.~\ref{fig:nc_film}, which was originally derived for doped STO films in low dielectric constant free of impurities environment. 
It is remarkable that due to the Rytova-Keldysh effect, relatively small total number of impurities of the film lead to the same disorder effect as much larger number of impurities in the insulating STO substrate. 

To emphasize importance of the Rytova-Keldysh effect, we briefly consider a case of doped STO film on relatively a thick buffer layer of undoped STO grown by MBE, which $\kappa_e = \kappa$ and concentration of impurities can be ignored. In this case, the amplitude of potential fluctuations~\footnote{In Eq.~\eqref{eq:gamma2} a logarithmic factor $\ln(\lambda/a)$ is neglected, so the results Eqs.~\eqref{eq:nc_normal1} and \eqref{eq:nc_normal2} are correct up to some logarithmic factors. }
\begin{equation}\label{eq:gamma2}
    \gamma \sim \frac{e^2}{\kappa \lambda} (Nd \lambda^2)^{1/2} = \frac{e^2}{\kappa} (Nd)^{1/2},
\end{equation}
and is independent on the screening length $\lambda$. 
Next we estimate the critical concentration $n_c$ using $\gamma \sim \mu$ similar to the previous section.
If $d \lesssim k_F^{-1}$, then $k_F \sim \sqrt{n d}$, and $\gamma \sim \mu$ gives
\begin{equation}\label{eq:nc_normal1}
    n_c \sim \frac{1}{a_B} \qty(\frac{N}{d})^{1/2}\qc d \lesssim d_2,
\end{equation}
where $d_2 = a_B/(N a_B^3)^{1/5}$ corresponds to $d \sim [k_F(n_c)]^{-1}$.
On the other hand, if $d \gtrsim k_F^{-1}$, then $k_F \sim n^{1/3}$, and $\gamma \sim \mu$ gives
\begin{equation}\label{eq:nc_normal2}
    n_c \sim \qty(\frac{N d}{a_B^2})^{3/4}\qc d_2 \lesssim d \lesssim d_c.
\end{equation}
At $d \gtrsim d_c$, $n_c$ crosses over to the bulk value Eq.~\eqref{eq:nc_bulk}.
For $N\simeq 10^{19}$ cm$^{-3}$ and $a_B \simeq 700$ nm, $d_2 \simeq 35$ nm. 
Equations.~\eqref{eq:nc_normal1} and \eqref{eq:nc_normal2} are shown by the red dot-dashed curve in Fig.~\ref{fig:nc_film}, which at $d \lesssim d_c$ is substantially lower than the blue curve. 
Thus we see that the Rytova-Keldysh effect allowing more distant impurities to contribute to potential fluctuations dramatically increases role of disorder. Similar enhancement of the role of Coulomb interaction in large dielectric constant films was studied for mobility~\cite{jena2007} and hopping conductivity~\cite{shklovskii2017}.

Theory of the percolation MIT developed in this section is valid for all other compensated semiconductor films, for example for PbTe films with thickness $d \lesssim d_c= 30$ nm. (For PbTe, the dielectric constant is $\sim 300$ with effective mass $\sim 0.2 m_e$ leading to $a_B \sim 80$ nm. Here following Ref.~\cite{Petrenko2014} we assume that PbTe film has $N \sim 10^{19}$ cm$^{-3}$ to estimate its corresponding $d_c = a_B/(N a_B^3)^{1/9} \simeq 30$ nm).

\appendix*
\section{Metal-insulator transition in STO wires}

\subsection{STO wire in low dielectric constant environment}

\begin{figure}[t]
    \centering
    \includegraphics[width = \linewidth]{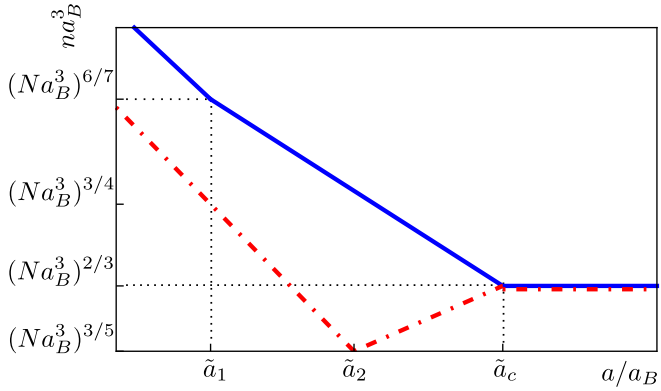}
    \caption{The phase diagram of MIT in the plane $(n,a)$, plotted in log-log scale using dimensionless variables $na_B^3$ and wire radius $\tilde{a} = a/a_B$. The blue curve corresponds to Eqs.~\eqref{eq:nc_wire1}, \eqref{eq:nc_wire2} and \eqref{eq:nc_bulk} in the case of doped STO wire both in low dielectric constant environment (say, silicon oxide) and on insulating STO substrate.
    The red dot-dashed curve corresponds to Eqs.~\eqref{eq:nc_wire_normal1}, \eqref{eq:nc_wire_normal2} and \eqref{eq:nc_bulk} in the case of doped STO wire on MBE STO substrate. Above each curve the system is metallic, while below the curve it is insulating. The dimensionless radius tick marks are $\tilde{a}_1 = (Na_B^3)^{-2/7}$, $\tilde{a}_2 = \tilde{d}_2$, and $\tilde{a}_c = \tilde{d}_c$.}
    \label{fig:nc_wire}
\end{figure}

In this subsection, we study critical electron concentration $n_c$ of metal-insulator transition in thin STO wires with dielectric constant $\kappa$ and radius $a$ in environment with dielectric constant $\kappa_e\ll\kappa$. In this case electrostatics shows~\cite{keldysh1997,finkelstein2002,teber2005,kamenev2006,cui2006} that all the electric field lines of a charged impurity channel through the thin wire before exiting outside to the environment at distance $\xi \simeq \sqrt{\kappa /2\kappa_e} a$.
The Coulomb potential of such a charge impurity at distance $a< \abs{x} < \xi \ln(\kappa/\kappa_e)$ decays exponentially $E_0 \xi e^{-\abs{x}/\xi}$ where $E_0 = 2e/\kappa a^2$.
Similar to the case of thin films, electrons screen this potential in wires at some screening radius $\lambda(n)$. 
At MIT, the screening becomes nonlinear and $\lambda$ can be estimated from the condition that fluctuations of impurity charge concentration in the volume $a^2 \lambda$ can be barely compensated by the redistribution of the concentration of electrons $n$, such that $n \sim (Na \lambda^2)^{1/2}/a^2\lambda$. This gives
\begin{equation}\label{eq:lambda_wire}
 \lambda \sim \frac{Na^2}{n^2}.
 \end{equation}
The corresponding amplitude of the Coulomb potential energy fluctuations is 
\begin{equation}\label{eq:gamma_wire}
    \gamma \sim e E_0 \lambda (Na^2 \lambda)^{1/2} \sim \frac{e^2 N^2 a^2}{\kappa n^3}.
\end{equation}
If $a \lesssim k_F^{-1}$, then $k_F(n) \sim na^2$, and $\gamma \sim \mu \sim h^2na^2/2m$ gives
\begin{equation}\label{eq:nc_wire1}
    n_c \sim \qty(\frac{N^2 }{a^8 a_B})^{1/5}\qc a \lesssim a_1,
\end{equation}
where $a_1 = a_B (N a_B^3)^{-2/7} \sim [k_F(n_c)]^{-1}$.
On the other hand, if $a \gtrsim k_F^{-1}$, then $k_F \sim n^{1/3}$, and $\gamma \sim \mu \sim h^2n^{2/3}/2m$ gives
\begin{equation}\label{eq:nc_wire2}
    n_c \sim \qty(\frac{N^2 }{a^4 a_B})^{3/11}\qc a_1 \lesssim a \lesssim a_c.
\end{equation}
where $a_c =a_B(Na_B^3)^{-1/9}=d_c$. At $a \gtrsim a_c$, the critical concentration of MIT, $n_c$, crosses over to the bulk value Eq.~\eqref{eq:nc_bulk}.
For $N\simeq 10^{19}$ cm$^{-3}$ and $a_B \simeq 700$ nm, we get $a_1 \simeq 10$ nm. 
Eqs.~\eqref{eq:nc_wire1} and \eqref{eq:nc_wire2} are shown by the blue curve in Fig.~\ref{fig:nc_wire}.
At $a \lesssim a_c$ it is higher than the blue curve in Fig.~\ref{fig:nc_wire}. 
This shows that STO wires are more vulnerable 
to impurities compared to STO films, because there are less electrons to screen the disorder potential fluctuations.

\subsection{Doped STO wire on the insulating STO substrate}
In order to emphasize the importance of the dielectric-constant-mismatch effect, we consider a case of doped STO wire on undoped STO grown by MBE, whose $\kappa_e = \kappa$ and concentration of impurities can be ignored. In this case, the amplitude of potential fluctuations 
\begin{equation}
    \gamma \sim \frac{e^2}{\kappa a} (Na^3)^{1/2},
\end{equation}
and is independent on the screening length $\lambda$.

If $a \lesssim k_F^{-1}$, then $k_F \sim na^2$, and $\gamma \sim \mu$ gives
\begin{equation}\label{eq:nc_wire_normal1}
    n_c \sim \qty(\frac{N}{a^7 a_B})^{1/4}\qc a \lesssim a_2,
\end{equation}
where $a_2 = a_B (N a_B^3)^{-1/5} \sim [k_F(n_c)]^{-1}$.
On the other hand, if $a \gtrsim k_F^{-1}$, then $k_F \sim n^{1/3}$, and $\gamma \sim \mu$ gives
\begin{equation}\label{eq:nc_wire_normal2}
    n_c \sim \qty(\frac{N a }{a_B^2})^{3/4}\qc a_2 \lesssim a \lesssim a_c.
\end{equation}
Notice Eq.~\eqref{eq:nc_wire_normal2} for a wire is identical to Eq.~\eqref{eq:nc_normal2} for a film, and $a_2 = d_2$~\footnote{This is because we ignore a logarithmic factor in Eq.~\eqref{eq:gamma2}.}.
At $a \gtrsim a_c=d_c$, the critical concentration of MIT $n_c$ crosses over to the bulk value Eq.~\eqref{eq:nc_bulk}.
For $N\simeq 10^{19}$ cm$^{-3}$ and $a_B \simeq 700$ nm, $a_2 \simeq 35$ nm. 
Equations.~\eqref{eq:nc_wire_normal1} and \eqref{eq:nc_wire_normal2} are shown by the red dot-dashed curve in Fig.~\ref{fig:nc_wire}. 
At $a \lesssim a_2$ it is substantially higher than the red dot-dashed curve in Fig.~\ref{fig:nc_film}.

Note that the above estimates of the critical concentration $n_c$ use the typical value $\gamma$ of random potential in the STO wire. Strictly speaking, in a wire with large length $L$ the chemical potential should exceed the largest potential barrier to make wire conducting. This adds factors proportional to $\ln (L/a)$ to our above estimates, which were skipped.

In the above discussion we assumed that there are no impurities outside the wire. 
Now we discuss the case with the same concentration of impurities $N$ in the STO substrate, but only wire has concentration $n$ of electrons. 
Then, a cube with side $R$ centered around a point of the wire creates a random charge of the order of $e\sqrt{NR^3}$. At large $R$ this charge is larger than 
total charge of electrons $enRa^2$ available for the cube screening. This means that the wire can not screen large scale and in the wire of length $L$ 
random potential fluctuations are of the size
\begin{align}
    \gamma \sim \frac{e^2}{\kappa} \frac{\sqrt{NL^3}}{L}.
\end{align}

Equating $\gamma$ and $\mu$, we obtain the critical concentration $n_c$ as follows
\begin{align}\label{eq:nc_wire_L1}
    n_c &\sim \qty(\frac{N L}{a_B^2 a^8})^{1/4}\qc a<a_3, \\
    n_c &\sim \qty(\frac{N L}{a_B^2})^{3/4}\qc a>a_3,\label{eq:nc_wire_L2}
\end{align}
where $a_3 = a_B (NL a_B^2)^{-1/4} \sim k_F(n_c)^{-1}$.
Eqs.~\eqref{eq:nc_wire_L1} and \eqref{eq:nc_wire_L2} are valid if $L>a_c$ and $L > a$. Eq.~\eqref{eq:nc_wire_L2} crosses over to the bulk value of $n_c$ Eq.~\eqref{eq:nc_bulk} at $L = a_c$.

\begin{acknowledgements}
We are grateful to J. Bharat, C. Leighton, D. Maslov, K.V. Reich, and B. Skinner for useful discussions. 
Y.H. was partially supported by the William I. Fine Theoretical Physics Institute.
\end{acknowledgements}

%

\end{document}